\documentclass{PoS}
\usepackage[authoryear,square]{natbib}
\bibpunct{(}{)}{;}{a}{}{,}

\title{The SKA and the Unknown Unknowns}

\ShortTitle{SKA and Unknown Unknowns}

\author{
\speaker{Peter Wilkinson}$^1$ 
\\ 
$^1$Jodrell Bank Centre for Astrophysics, School of Physics and Astronomy, The University of Manchester, Manchester, UK;\\

\\
\\
E-mail: \email{peter.wilkinson@manchester.ac.uk}
}

\abstract{As new scientists and engineers join the SKA project and as
  the pressures come on to maintain costs within a chosen envelope it
  is worth restating and updating the rationale for the ``Exploration
  of the Unknown''~(EoU).  Maintaining an EoU philosophy will prove a
  vital ingredient for realizing the SKA's discovery potential.  Since
  people make the discoveries enabled by technology a further axis in
  capability parameter space, the ``human bandwidth'' is emphasised.
  Using the morphological approach pioneered by Zwicky, a currently
  unexploited region of observational parameter space can be
  identified viz: time variable spectral patterns on all spectral and
  angular scales -- one interesting example would be ``spectral
  transients''.  We should be prepared to build up to 10\% less
  collecting area for a given overall budget in order to enhance the
  ways in which SKA1 can be flexibly utilized.  }

\FullConference{
Advancing Astrophysics with the Square Kilometre Array\\
June 8-13, 2014\\
Giardini Naxos, Sicily, Italy}

\newcommand{\skipthis}[1]{}

\begin{document}

\section{Introduction}\label{intro}
The 2014 SKA Science Case has become impressively wide-ranging.
Perhaps the most significant change in perception over recent years
has been the increased importance afforded to transient science, which
can be viewed as an emerging response to the ``6th KSP'' the
Exploration of the Unknown (EoU).  Nonetheless the main science
imperatives for SKA1 can be traced back to the 2004 Science
Case~\citep{C04} -- and quite often back to the arguments for the
``Hydrogen Array''~\citep{W91}.  The rationale for the SKA has
firm foundations.

Despite the growth in breadth and depth of the science case over the
past ten years we can still be certain that the SKA's place in
astronomical history will not all be found within its pages. As the
authors of EoU \citep{W04} in the 2004 Science Case put it ``If
history is any example, the excitement of the SKA will not be in the
old questions which are answered, but the new questions which will be
raised by the new types of observations it alone will permit. The SKA
is a tool for as-yet-unborn users and there is an onus on its
designers to allow for the exploration of the unknown.'' More
recently~\citet{K12} has noted: ``discovery is important in astronomy
because we are not sufficiently imaginative to construct the Universe
and its constituents from first principles...'' 

The question is how to respond.  In 2004 when considering the
potential Key Science Programmes the SKA Level Zero Science Committee
stated: ``... the topic of `serendipity' does not meet the [level-0]
definitions... The recommendation... is therefore that serendipitous
discoveries and the expansion of phase space not be included as
level-0 science but that serendipity be explicitly included in the
science case as an additional motivation for building the SKA.''  As
Carilli put it in his science retrospective to AASKA14 ``unknowns were
a hard sell in 2004 since physics had torqued astronomy towards
experiments but the pendulum is swinging... and general facilities may
be back in vogue''.

Recent SKA planning documents show that indeed the arc of history is
bending in the right direction.  In the SKA Concept of Operations
Document~\citep{B13} for example: ``the primary success metric for the
SKA Observatory will be the significance of its role in making
fundamental scientific discoveries'' and ``recognizing the long
history of discovery at radio wavelengths... the telescope will be
designed in a manner to affordably allow flexibility and evolution of
its capabilities to probe new parameter space (e.g., time variable
phenomena that current telescopes are not well-equipped to
detect)''. While in the SKA Level 0 Science Requirements
Document~\citep{B14} three of the general scientific requirements of
SKA1 are given as:
\begin{itemize}
\item to provide the capability to carry out small-scale,
experimental observations that utilise the outputs of tied-array beams
from one or more sub-arrays.
\item to be capable of rapid reconfiguration
of their observing mode in response to internal or externally
generated, pre-defined triggers. 
\item to enable commensal
observations. It is apparent that many scientific applications can
make effective use of the same basic observing campaign to multiply
the scientific productivity of the facilities by a large factor.
\end{itemize}
Explicit realizations of these philosophies are captured in the
specifications for the SKA1 Level 0 science goals~\citep{C14} . This
is all greatly to be welcomed.  But as pressures come on to keep costs
within a chosen ceiling good intentions may fall victim to bean
counting.

\section{Discovery in Radio Astronomy -- Lessons from History}

In 2007 I assembled a personal list of important discoveries made with
classical radio astronomy instruments working from deca to centimetre
wavelengths -- as is appropriate for SKA \citep{W07}. Space
prevents me from presenting a complete updated list here, instead
additions to the 2007 list (with thanks to colleagues) are given in
Table 1. What are the lessons to be learned ?

\begin{enumerate}

\item In this region of the spectrum the large telescopes of their day
  have been at the forefront of discovery. This can be ascribed to the
  inherently low signal-to-noise ratio of observations indeed it is
  only with a radio telescope the size of the SKA that one can expect
  to detect the radio emission from most objects detected in other
  wavebands.

\item The rate of discovery has slowed in the last ten years (see
  Table 2) despite the rapid improvement in digital technology which
  has been well harnessed by the pulsar community.This was the period
  when LOFAR was being constructed and the VLA was being upgraded to
  the JVLA so we can hope to recover from this discovery ``recession''
  over the course of this decade. But to me Table 2 is a prima facie
  argument for the SKA.

\item The designers of did not anticipate what they would be ``known
  for'' -- a failing shared with HST as pointed out by \citet{N15}.

\item In many cases, the people involved had access to lots of
  telescope time and were not testing any particular theoretical model
  (not evidenced in the Table).

\end{enumerate}
\begin{table}
\begin{tabular}{|p{4.5cm}|p{3.5cm}|p{4cm}|p{2cm}|}
\hline {\bf Discovery}&{\bf Instrument}&{\bf Design Purpose}&{\bf
  Forseen by Designers?}\\ \hline

Interplanetary Scintillation (1955)&Special purpose
interferometer&Solar radio astronomy&Partly\\\hline

Ordered Magnetic Fields in External Galaxies (1970s)&Westerbork SRT
and Effelsberg&High resolution imaging General purposes&No\\\hline

Laing-Garrington effect (1988)&Cambridge 5km and Very Large array&High
resolution imaging&No\\\hline

IDV is interstellar scintillation (1992)&Very Large Array Westerbork
SRT&High resolution imaging &No\\\hline

Spinning Dust (1996/97)&COBE and OVRO 40m&CMB and general
purpose&No\\\hline

Geodetic (GR) Precession in a binary pulsar (1998)&Effelsberg&General
purpose&No\\\hline

Winds on Titan (2004)&Global VLBI&Mapping compact sources&No\\\hline

RRATs--intermittent pulsars (2006)&Jodrell Bank Lovell
Telescope&General Purpose&No\\\hline

Fast Radio Bursts (2006)&Parkes&General purposes&No\\\hline

Neutron stars above 2Msun (2010)&GBT&General purposes&No\\\hline
\end{tabular}
\caption{Discoveries in deca-centimetric radio astronomy -- an update to the list in Wilkinson (2007)}
\end{table}

\begin{table}
\begin{center}
\begin{tabular}{|l|c|}
\hline
{\bf Decade} &{\bf Number of discoveries} \\
\hline
1930-9&1\\
1940-9&2\\
1950-9&4\\
1960-9&8\\
1970-9&6\\
1980-9&5\\
1990-9&8\\
2000-9&4\\
\hline
\end{tabular}
\end{center}
\caption{Rate of discovery in the deca-centimetric radio band}
\end{table}

\section{Discovery with SKA}

\subsection{Parameter Space }
          
The SKA's place in astronomical history will not all be anticipated
within the pages of the 2014 Science Case -- so what do we do about
it?  We start by recognizing that discovery can be planned. As
\citet{H84} noted technology innovation allows the exploration of new
areas of capability parameter space whose axes are, broadly:
\begin{itemize}
\item	sensitivity; 
\item	spatial coverage and or resolution; 
\item	temporal coverage and or resolution;
\item	spectral coverage and or resolution;
\item	availability of past observations.
\end{itemize}

Note that the last one is already an addition to Harwit's original
thinking. The ready access to archive data has become a major factor
in many modern scientific projects.  A recent highlight in radio
astronomy was the discovery of Fast Radio Bursts (FRBs) in the Parkes
pulsar archive some years after the data were taken \citep{L07}.
Other pulsar discoveries are mentioned below.

In terms of these technologically-driven axes the future looks
bright. Even SKA1 will offer a huge boost in sensitivity coupled with
a wide field of view for survey throughput plus the ability to sample
with high resolutions in the time, frequency and spatial domains.
Flexible archiving is also seen as a priority.  Thus in terms of
technical capability we should expect SKA1 to be able to find
unexpected phenomena-- but only if ``the system'' allows its users the
maximum chance to do so.  To encapsulate this idea let us identify a
new, more abstract, axis in capability parameter space

\begin{itemize}

\item human bandwidth.  
\end{itemize}

Technological innovation alone will not be enough to unlock SKA's full
discovery potential.  There should be scope for many as many people as
possible to gain access to SKA data and for the expression of
curiosity in its use.

\subsection{Types of discovery}

In the International SKA SWG \citep{C06} put it succinctly:
``One might think that all of the combinations of key variables have
been probed with existing telescopes. This is far from being the case
because the so-defined parameter space has been investigated only in
very compartmentalized sub-volumes.''  They went on to extend the
rather general arguments adduced by \citet{W04} and
introduced a wide range of potential areas for discovery: 
\begin{enumerate}

\item New
objects in known classes.  

\item Targeted known phenomena e.g.  
\begin{itemize}

\item the EoR;

\item magnetic fields in structures covering a range of physical
scales 

\item coherent emission from extrasolar planets 

\end{itemize}

\item New Classes of Sources and Phenomena Based on Known Physics,
  Biology e.g.

\begin{itemize}

\item transient sources of many kinds 

\item new structures e.g in the Cosmic Web

\end{itemize}

\item The totally unexpected

\end{enumerate}
 For the ``rare events'' under heading 1) the path forward is simple -
 find more objects. The amount of ``surprise'' in a data set rises
 only as the logarithm of the number of independent data elements
 within it.  In the context of astronomical discovery this
 ``logarithmic surprise'' idea was first used by
 \citet{D98}. \citet{W07} particularized it for pulsar-related
 discoveries and the pulsar story was extended by \citet{K14}.  For
 the targeted known phenomena under heading 2) the way forward is
 basically via the KSPs and PI science -- after all these are the
 ``known unknowns''.  However when we come to headings 3) and 4) new
 classes of sources and ``allowed'' phenomena and the complete
 surprises we need to look in regions of parameter space not
 previously explored.  As \citet{C06} recognized, the problem
 arises in defining these areas and so we now turn to this issue.

\subsection{Are there unexplored regions of parameter space?} 

As outlined by \citet{W07} a purely phenomenological approach can
offers a complement to the thinking described by \citet{C06}
and by \citet{C07}.  This is the basis of Zwicky's ``morphological''
methodology which he developed to a high art in astronomy and other
fields (e.g. \citet{Z69}). Its basis is that all volume elements in
parameter space - as long as they are not explicitly prohibited --
should be considered.  Identifying these elements involves freeing
one's mind of prejudice and being willing to contemplate unorthodox
comparisons.  A start for assessing the reach of SKA1 Level 0 science
can be made by constructing a cross-linking matrix with the following
broad-brush parameters: 
\begin{itemize}

\item Observing time: short \& long 

\item Total intensity: high \& low SNR

\item Polarised intensity: high \& low SNR 

\item Spectral patterns: small scale \& large scale 

\item Spectral resolution: high \& low

\item Spatial patterns: small scale \& large scale 

\item Spatial resolution: high \& low

\item Temporal patterns: short term \& long term 

\item Temporal resolution: high \& low 
\end{itemize}

This leads to a matrix with 18 rows and columns and 144 independent
boxes.  This matrix is not shown here for reasons of space but the
SKA1 Level 0 Science programmes \citep{C14} do a good job in
filling the boxes.  There is, however, a set of boxes involving time
variable spectra on all angular and spectral scales which is not
filled. One example would be ``spectral transients''.

\subsection{Operational flexibility}

In order to explore observational parameter space as widely as
possible some broad flexibility principles were laid out by
\citet{W04}. It is encouraging that many of these principles are
represented, in one form or another, in the Level 0 Science
Specifications and the steadily solidifying technical design.
However, the cardinal point of this paper is that people make
discoveries not technology. The problem is that no one can forecast
which right people are going to be in the right place at the right
time. The only response must be to seek ways to allow some people
access to lots of telescope time and to allow as many different people
as possible to gain some access to the data.  The essence of the human
bandwidth axis is maximising the ``brain multiplexing factor''.  With
this in mind let us look at the range of operational modes identified
in SKA documents \citep{B13,B14,C14}.  Reduced to their simplest they
are:

\begin{itemize}
\item	Normal
\begin{itemize}
               \item PI
               \item Legacy survey
\end{itemize}
\item	Targets of Opportunity
\item	Custom Experiments 
\item	Commensal Observing
\end{itemize}
A few comments seem worthwhile.  First, the normal and TOO modes will
be evolutions of existing practice but there is always scope for
adding flexibility.  For example when the data are being taken for the
Legacy Surveys can one open the ``collection window'' more than is
strictly necessary so as to add depth to the archived data set and
hence increase the post-observation discovery potential?  Second, the
most radical of the suggestions made by \citet{W04} was to allow a
fraction of observing time for high-risk or unproven new-style
observations which may be tentative or based on a hunch and hence hard
to justify in the formal sense. Here the availability of independent
beams and commensal observing will help without compromising
conventional observing programs. Custom experiments must be considered
on a case-by-case basis -- but don't become prey to the criticism ``
knows the price of everything and the value of nothing''.  Finally, a
constant theme of the SKA should be to seek constructive ways of
establishing parallel access to the data.  Wider public access is only
briefly mentioned in \citet{B13} but it could well be crucial in EoU.
``Citizen Science'' is a proven route to maximising the ``reach'' and
the recent run of successes of Einstein@home in finding new pulsars
\citep{A13,K13,P13} provides graphic evidence.  The SKA was born
global -- so let it be a globally accessible source of data!

\subsection{Synergy with other telescopes}
The sensitivity of the SKA means that for the first time in the
history of astronomy, an object detected in any waveband in the
electromagnetic spectrum is also likely to be detected in the radio.
The opportunities for data linkages with other with ground-
(e.g. LSST; ELT; TMT) and space-based (e.g. JWST; GAIA; EUCLID)
telescopes will be enormously greater than with any previous radio
telescope.  As well as specific bilateral comparisons SKA scientists
should make a major commitment to getting the most out international
Virtual Observatories and hence to maximise professional and public
access to multi-wavelength data.  There was plenty of evidence from
talks in in AASKA14 that this message is understood by the SKA
community.

\subsection{Conclusions}
\begin{center}
``The known knowns'' -- read the literature;\\
``The known unknowns'' -- write a proposal;\\
``The unknown unknowns'' -- hold on to a vision.\\
\end{center}

The current community which is planning and implementing the first
phase of the SKA should maintain faith with the scientists of the
future by ensuring that the SKA system is prepared for the EoU.  This
capability will come from a holistic vision of its: 
\begin{itemize}
\item Large data
collecting power: from raw sensitivity and large field-of-view plus
multiple beams some of which can be dedicated to specific tasks.  
\item Some freedom to take risks: from use of independent beams plus new
time allocation paradigms.  
\item Large human bandwidth: allowed by
commensal observing; multiple independent beam capability; an in depth
data archive coupled with a coordinated strategy for the use of
international Virtual Observatories and the citizens of the world.
\end{itemize}
Much is this is there in present thinking.  But the system is not yet
built and the flexibility required for the EoU is likely to impose an
additional cost burden and hence is vulnerable.  What price is
reasonable? I propose that we should be prepared to build up to 10\%
less collecting area for a given overall budget in order to enhance
the ways in which this area can be utilized. The penalty may be up to
20\% longer to carry out some projects but this is a graceful
degradation and the upside is opening up ``opportunity space''.  I
believe that this is likely to prove a good bargain.

\bibliographystyle{apj}

\end{document}